\documentclass[aps,showpacs,nofootinbib]{revtex4}
\usepackage{epsf,latexsym}
\usepackage{amsfonts}

\begin{document}

\title{Spacetime atoms and extrinsic curvature 
of equi-geodesic surfaces}
\author{Alessandro Pesci}
\email{pesci@bo.infn.it}
\affiliation
{INFN-Bologna, Via Irnerio 46, I-40126 Bologna, Italy}

\begin{abstract}
A recently-introduced function $\rho$
of spacetime event $P$
expressing spacetime  
as made of `spacetime atoms'
of quantum origin
is considered. 
Using its defining relation,
we provide an exact expression for $\rho$ 
involving the van Vleck biscalar, 
and show it can be recast 
in terms of the extrinsic curvature of suitable equi-geodesic surfaces
centered at $P$.
Moreover,
looking at the role $\rho$ plays
in the statistical description of spacetime,
we point out that this quantity should actually be 
understood
as counting the quantum states 
of the collection of spacetime atoms
rather than counting directly the spacetime atoms themselves
(or the degrees of freedom associated to them),
and would correspond to the ratio of  
the number of quantum states `at $P$'
for an assigned spacetime configuration
to the number of quantum states
for flat spacetime. 
\end{abstract}


\maketitle

$ $

In \cite{PadX, PadY, PadZ, Pad00},
a statistical description of spacetime
in terms of a function of the number density 
of `spacetime atoms'
has been introduced.
The context in which this notion arises, 
is a description of any $D$-dimensional spacetime,
first given in \cite{KotD},
in terms of a sort of effective `metric' $q_{ab}$
(quasimetric, qmetric),
which conforms to 
the geodesic structure of the manifold
and provides
a squared geodesic distance between two events
which asymptotically coincides with that given by
the usual metric $g_{ab}$ when the events are `far apart',
but gives $\pm L^2$ 
(depending the sign on the spacelike or timelike
nature of the displacement) 
in the coincidence limit,
where $L$ is a minimal length scale.

The discreteness of spacetime
at event $P$ based on minimal length $L$,
is meant to be captured by a function 
$\rho = \rho(P, n^a)$, 
related to the number density of `atoms of spacetime'
at $P$ 
in the direction given by any unit 
$D$-dimensional vector $n^a$,
defined
on the space corresponding to the Euclideanisation of a given spacetime
at $P$.

In \cite{PadX, PadY, PadZ, Pad00},
it is noticed that 
$\rho$, 
defined in more detail as we are going to describe in a moment, 
can be written as
\begin{eqnarray}\label{pad}
\rho(P, n^a) 
&=&
1 - \frac{1}{6}  L^2 E + o(L^2 E),
\end{eqnarray}
where $R_{ab}$ is the Ricci tensor
and $E := R_{ab} \ n^a n^b$.
%
%
Our aim here, 
is to investigate
a possibly exact 
expression for $\rho$,
giving of course
equation (\ref{pad})
in the suitable limit,
envisaged already
in \cite{PadX} itself,
and try to elaborate on the
interpretation of  $\rho$. 

Let us recall how $\rho$ is introduced.
In doing this,
we choose to remain in the Lorentz sector,
i.e. to have no reliance on the Euclideanised space.
Let $P$ be a point in $D$-dimensional spacetime.
The congruence of all timelike geodesics emanating from $P$ should contain
all what is needed to describe the spacetime in a neighbourhood of $P$,
since it accounts for
all possible observers freely falling through $P$. 
The first object we define is the hypersurface
$\Sigma(P, l)$ of all points $p$
at assigned distance along timelike geodesics from $P$:

\begin{eqnarray}
  \Sigma(P, l) = \big\{p: \ \sigma^2(p, P) = -l^2 \ (<0)\big\},
  \nonumber
\end{eqnarray}
where 
$
\sigma^2(p, P)
$
is the squared geodesic distance between $P$ and $p$
($\sigma^2(p, P) = 2 \Omega(p, P)$, where $\Omega(p, P)$ is the Synge
world function biscalar \cite{Syn}),
and we are taking $l = \sqrt{l^2}$ non-negative.

The Euclidean definition of $\rho$
is given in terms of (generic/flat ratio of) areas
of hypersurfaces formed by points
at an assigned geodesic Euclidean distance $d$ from $P$
in the limit $d \rightarrow 0$,
the fundamental feature being that,
according to the effective metric,
these $(D-1)$-areas remain finite
in the coincidence limit in which the hypersurfaces shrink to $P$ \cite{PadW}
(and clearly, one would expect some analogous results
do hold true in the Lorentzian sector). 
We can proceed analogously here along the following lines.
For each timelike normalised vector $n^a$ at $P$,
we consider the intersection point $\hat p$ between
the timelike geodesic $\gamma(n^a)$
with tangent $t^a(P) = n^a$ at $P$
and the hypersurface $\Sigma(P, l)$.
Let $y^i$, $i = 1, ..., D-1$
be coordinates on $\Sigma(P, l)$ 
such that $y^i({\hat p}) =0$.
We consider a segment $I$ of hypersurface $\Sigma(P, l)$
around $\hat p$, defined as
$
I = \{dy^i\},
$
where $dy^i$ are thought of as fixed when $l$ is varied.
The $(D-1)$-dimensional area of $I$ is

\begin{eqnarray}
  d^{D-1}V({\hat p}) = \sqrt{h({\hat p})} \ d^{D-1}y, \nonumber
\end{eqnarray}
where $h_{ij}$ are the components of the (spatial) metric
on $\Sigma(P, l)$
in the coordinates $y^i$,
metric which coincides with that induced
by the spacetime metric $g_{ab}$.
What we have to consider is the area $[d^{D-1}V]_q$ of $I$ 
as measured through the effective metric $q_{ab}$.

We know
the effective metric at any point $p$ in a neighbourhood of $P$
is described in terms of a bitensor
$q_{ab}(p, P)$ \cite{StaA}
which stems from requiring that the squared
geodesic distance gets modified into
$
\sigma^2 \rightarrow [\sigma^2]_q =
      {\tilde S_L}(\sigma^2) = S_L(p, P) 
$
with
{\bf (R1)} $S_0 = \sigma^2$,
{\bf (R2)} ${\tilde S_L}(0^\pm) = \pm L^2$,
${\tilde S}^\prime(0)$ finite
($^\prime$ indicates differentiation with respect
to its argument $\sigma^2$),
and
{\bf (R3)}
the kernel $G(\sigma^2)$ of the d'Alembertian
gets modified into
$G(\sigma^2) \rightarrow [G]_q(\sigma^2) = G({\tilde S_L})$
in all maximally symmetric spacetimes.
These requirements give, for spacelike or timelike geodesics,
the expression

\begin{eqnarray}\label{qab}
  q_{ab}(p, P) = A(p, P) g_{ab}(p) +
  \epsilon \Big(\frac{1}{\alpha(\sigma^2)} - A(p, P)\Big) t_a(p) t_b(p),
\end{eqnarray}
where $t^a$ is the tangent vector with the normalization
$
g_{ab} t^a t^b = \epsilon = \pm 1,
$

\begin{eqnarray}\label{A}
  A(p, P) = \frac{S_L(p, P)}{\sigma^2(p, P)}
  \Big(\frac{\Delta(p, P)}{\Delta_S(p, P)}\Big)^\frac{2}{D-1},
\end{eqnarray}

\begin{eqnarray}\label{alpha}
  \alpha = \frac{\tilde S_L}{\sigma^2 ({\tilde S}^\prime_L)^2}.
\end{eqnarray}
Here

\begin{eqnarray}\label{vanVleck}
  \Delta(p, P) = - \frac{1}{\sqrt{g(p) g(P)}}
  {\rm det}\Big[-\nabla^{(p)}_a \nabla^{(P)}_b \frac{1}{2} \sigma^2(p, P)\Big]
\end{eqnarray}
is the van Vleck determinant \cite{Xen,PPV} which is a biscalar,
and the biscalar $\Delta_S(p, P)$ is
$\Delta_S(p, P) = \Delta({\tilde p}, P)$,
where $\tilde p$ is
that point on the geodesic through $P$ and $p$
(on the same side of $p$ with respect to $P$)
which has $\sigma^2({\tilde p}, P) = S_L(p, P)$.
From
the effective metric $[h_{ab}]_q(p, P)$ induced by $q_{ab}(p, P)$
at points $p$ on a hypersurface
at $\sigma^2(p, P) = {\rm const}$,
we get 
the effective-metric $(D-1)$-dimensional area of $I$ as
\begin{eqnarray}
 [d^{D-1}V]_q({\hat p}, P)
 = \Big[\sqrt{h}\Big]_q({\hat p}, P) \ d^{D-1}y . \nonumber
\end{eqnarray}

In analogy with the Euclidean version,
$\rho$ can then be defined considering the ratio
of the effective-metric $(D-1)$-dimensional area of $I$
for the actual configuration, $[d^{D-1}V]_{q(g)}({\hat p}, P)$, 
to what we would have were the spacetime flat, $[d^{D-1}V]_{q(\eta)}({\hat p}, P)$
($\eta_{ab}$ is Minkowski metric), in the
limit ${\hat p} \rightarrow P$ along $\gamma(n^a)$,
i.e.

\begin{eqnarray}\label{rho_def}
  \rho(P, n^a) =
  \bigg(\lim_{{\hat p} \rightarrow P}
  \frac{[d^{D-1}V]_{q(g)}({\hat p}, P)}{[d^{D-1}V]_{q(\eta)}({\hat p}, P)}\bigg)
  _{\gamma(n^a)} .
\end{eqnarray}

Our aim is now to express $\rho$
in terms of the quantities $A$ and $\alpha$ defining the effective metric.
The effective metric $[h_{ab}]_q$ induced by $q_{ab}$
on a hypersurface
at $\sigma^2 = {\rm const}$ 
turns out to be

\begin{eqnarray}
  [h_{ab}]_q(p, P) = A(p, P) h_{ab}(p)
\end{eqnarray}  
\cite{KotF},
which implies

\begin{eqnarray}
  \Big[\sqrt{h}\Big]_q(p, P) = A(p, P)^{\frac{D-1}{2}} \sqrt{h(p)}.
\end{eqnarray}  
From this,
the effective-metric $(D-1)$-dimensional area of $I$
is found to be

\begin{eqnarray}
  [d^{D-1}V]_q({\hat p}, P)
  &=& \Big[\sqrt{h}\Big]_q({\hat p}, P) \ d^{D-1}y \nonumber \\
  &=& A({\hat p}, P)^{\frac{D-1}{2}} \sqrt{h({\hat p})} \ d^{D-1}y \nonumber \\
  &=& A({\hat p}, P)^{\frac{D-1}{2}} d^{D-1}V({\hat p}), \nonumber
\end{eqnarray}
where $d^{D-1}V({\hat p})$ indicates the proper area of $I$
according to the ordinary metric.
Here we see that
only $A$, and not $\alpha$,
is involved
in the determination of $\rho$.

For each given timelike versor $n^a$ at $P$,
we choose a local Lorentz frame at $P$
such that  
$
n^a \buildrel{*}\over= \delta^a_0
$
($\buildrel{*}\over=$ indicates that the r.h.s.
is what the expression in the l.h.s. is
in the specified frame)
and introduce an angular coordinate $\eta$ on $\Sigma(P, l)$ 
in a neighbourhood of $\hat p$,
defined by requiring
that $l d\eta$ be proper length
on $\Sigma(P, l)$,
and that $\eta({\hat p}) = 0$.
We consider $(D-1)$ mutually orthogonal coordinates $z^i$ on $\Sigma(P, l)$
such that $z^i = l \eta$,
and choose as $I$ the (hyper)cube ${dz^i}$
at $\hat p$ defined by $dz^i = l d\eta, \forall i$.
This gives

\begin{eqnarray}
  [d^{D-1}V]_q({\hat p}, P)
  &=& A({\hat p}, P)^{\frac{D-1}{2}} l^{D-1}
      \big(1 + {\cal O}(l^2)\big) (d\eta)^{D-1}
      \nonumber
\end{eqnarray}
where the ${\cal O}(l^2)$ term
represents the effects of curvature
(and is thus of course absent in the flat case),
and clearly
$l = l({\hat p}, P) = 
\sqrt{-\sigma^2({\hat p}, P}).
$
Using the expression (\ref{A}) for $A$,
we get

\begin{eqnarray}
  [d^{D-1}V]_q({\hat p}, P)
  &=& \bigg[\frac{\epsilon S_L({\hat p}, P)}{l^2}\bigg]^{\frac{D-1}{2}}
      \frac{\Delta({\hat p}, P)}{\Delta_S({\hat p}, P)} \
      l^{D-1} \big(1 + {\cal O}(l^2)\big) (d\eta)^{D-1}   \nonumber \\
  &=& [\epsilon S_L({\hat p}, P)]^{\frac{D-1}{2}}
      \frac{\Delta({\hat p}, P)}{\Delta_S({\hat p}, P)} \
      \big(1 + {\cal O}(l^2)\big) (d\eta)^{D-1}.   \nonumber
\end{eqnarray}
Taking the limit ${\hat p} \rightarrow P$ along $\gamma(n^a)$,
we have

\begin{eqnarray}
  \lim_{{\hat p} \buildrel{\gamma(n^a)}\over\longrightarrow P}
  [d^{D-1}V]_q({\hat p}, P)
  = L^{D-1} \frac{1}{\Delta_L(P, n^a)} (d\eta)^{D-1},     \nonumber
\end{eqnarray}  
where
$\Delta_L(P, n^a) = \Delta({\bar p}, P)$, where $\bar p$ 
is that point on geodesic $\gamma(n^a)$ (on the side in the direction $n^a$)
which has $\epsilon \sigma^2({\bar p}, P) = L^2$.

This shows that the numerator and the denominator
in expression (\ref{rho_def}) remain both non vanishing
in the coincidence limit ${\hat p} \rightarrow P$,
exactly as it happens in the Euclidean case.
Indicating with
$[d^{D-1}V]_q(P, n^a)$ this limit for the actual configuration
and with $[d^{D-1}V]_q(P, n^a)_{flat}$ the limit one would obtain
were the spacetime flat,
we have

\begin{eqnarray}
 [d^{D-1}V]_q(P, n^a) =  L^{D-1} \frac{1}{\Delta_L(P, n^a)} (d\eta)^{D-1}
 \nonumber
\end{eqnarray}
and,
since for flat spacetime $\Delta(p, P) = 1$ identically and then
$\Delta_L(P, n^a) = 1$,

\begin{eqnarray}
  [d^{D-1}V]_q(P, n^a)_{flat} = L^{D-1} (d\eta)^{D-1}.   \nonumber
\end{eqnarray}
We get thus

\begin{eqnarray}\label{rho}
  \rho(P, n^a)
  &=& \frac{[d^{D-1}V]_q(P, n^a)}{[d^{D-1}V]_q(P, n^a)_{flat}} \nonumber \\
  &=& \frac{1}{\Delta_L(P, n^a)}.
\end{eqnarray}
%

This is an exact formula for $\rho$ in terms of the
van Vleck determinant 
(along with what envisaged in \cite{PadX}),
with the latter given by expression (\ref{vanVleck})
evaluated at $\epsilon \sigma^2 = L^2$ along $\gamma(n^a)$.
There is, mathematically, nothing special in our choice
of timelike geodesics instead of spacelike geodesics in spacetime
or, also, of geodesics in a Riemannian space (such as that supposed to come
out from the Euclideanisation procedure at $P$), and the expression for
$\rho(P, n^a)$ is identical in these cases, with spacelike $n^a$
and with $\Delta(p, P)$ the van Vleck determinant of the spacetime/space
under consideration. 
An expansion of $\Delta(p, P)$ in powers of $l = \sqrt{\epsilon \sigma^2}$
can be given (cf. \cite{Xen}),  
with coefficients given by scalars constructed from
$E_{ab} \equiv R_{acbd} t^c t^d$, and their derivatives of any order.
At leading order, this reads

\begin{eqnarray}\label{Delta_expansion}
  \Delta(p, P) = 1 + \frac{1}{6} l^2 E(P) + o\big(l^2 E(P)\big),
\end{eqnarray}  
with
$
E = g^{ab} E_{ab} = R_{ab} t^a t^b,
$
where $t^a$ is normalised timelike/spacelike.
This gives

\begin{eqnarray}\label{DeltaL}
 \Delta_L(P, n^a) =  
 1 + \frac{1}{6} L^2 R_{ab}(P) n^a n^b + o\big(L^2 R_{ab}(P) n^a n^b\big),
\nonumber
\end{eqnarray}
in equation (\ref{rho}),
and
we get expression
(\ref{pad}).

$\Delta$ can be expressed
in closed form 
in terms 
of the extrinsic curvature $K$
of the surface $\Sigma$.
In fact,
from the relation \cite{StaA}

\begin{eqnarray}\label{S&K}
\frac{\partial}{\partial l} \ln \Delta 
&=& \frac{D-1}{l} - K(P, l), 
\end{eqnarray}
where $K(P, l)= K(P, l({\hat p}, P))$ is the extrinsic curvature,
regarded as a function of $P$ and $l$,
of the surface $\Sigma(P, l)$,
we get

\begin{eqnarray}\label{ln_vV}
\ln \Delta 
&=& \int_0^l
\Big[\frac{D-1}{\tilde l} - K(P, {\tilde l})\Big] 
d{\tilde l} + C \nonumber \\
&=& \int_0^l
\Big[K^{flat}(P, {\tilde l}) - K(P, {\tilde l})\Big]
d{\tilde l} 
\end{eqnarray}
where the integrand is necessarily finite
in the limit ${\tilde l} \rightarrow 0$
(in fact, it goes to 0 \cite{KotF}),
and the integration constant $C$ must be 0
in view of $\Delta(0) = 1$ \cite{Xen, PPV}.
Also, as in \cite{KotF},
we explicitly use of that
$\frac{D-1}{l}$  is
the extrinsic curvature
$K^{flat}(P, l)$ 
of $\Sigma(P, l)$ 
in flat case. 
Then,

\begin{eqnarray}
\Delta 
&=& 
e^{\int_{0}^l 
[K^{flat}(P, {\tilde l}) - K(P, {\tilde l})] 
d{\tilde l} }   \nonumber \\
&=&
e^{\mu({\hat p}, P)},
\end{eqnarray}
where $\mu({\hat p}, P)$ denotes the integral in the exponent.

Thus,

\begin{eqnarray}
\Delta_L(P, n^a) 
&=&
\Delta({\bar p}, P)    \nonumber \\
&=&
e^{\mu_L(P, n^a)},
\end{eqnarray}
where
$\bar p$ is along $\gamma(n^a)$ at distance
$l = L$ and
$\mu_L(P, n^a) = \mu({\bar p}, P)$,
i.e.

\begin{eqnarray}\label{mu_L}
\mu_L(P, n^a) 
&=&
\int_{\gamma_L(n^a)}(K^{flat}-K),
\end{eqnarray}
where $\gamma_L(n^a)$ is the geodesic segment
of proper length $L$ along $\gamma(n^a)$
and 
$(K^{flat}-K)$ is the 1-form
$[K^{flat}(P, l)-K(P, l)] dl$.
And, in term of this,

\begin{eqnarray}\label{f_2}
\rho(P, n^a) =
e^{-\mu_L(P, n^a)}.
\end{eqnarray}
%
%
This is again an exact formula for $\rho$.
It is equation (\ref{rho}) expressed in terms
of extrinsic curvature.

Since \cite{KotF}

\begin{eqnarray}
K^{flat}(P, l) - K(P, l) =
\frac{l}{3} \ E(P) + o(\sigma \ E(P)),   
\end{eqnarray}
equation (\ref{mu_L}) gives

\begin{eqnarray}\label{mu_L_2}
\mu_{L}(P, n^a) =
\frac{1}{6} L^2 E(P) + o(L^2 E(P)). 
\end{eqnarray}

Using this,
we can write
the expression $e^S$ 
given in \cite{Pad00}
for the total number of states
of matter (with energy momentum tensor $T_{ab}$)
and quantum spacetime
in a region ${\cal R}$ of
$D=4$ spacetime
as (we are using $\hbar = c = k_{B} = 1$ units)

\begin{eqnarray}\label{expS}
e^S
&\propto& 
\prod_{n} \ \prod_{x} \ \rho \ \rho_m     \nonumber \\
&=&
\prod_{n} \ \prod_{x} \ \rho \ e^{L^4 T_{ab} n^a n^b}  \nonumber \\
&=&
\prod_{n} \ \prod_{x} \ e^{-\mu_L + L^4 T_{ab} n^a n^b}    \nonumber \\ 
&\approx&
\prod_{n} \ \prod_{x} \  e^{-\frac{1}{6} L^2 E + L^4 T_{ab} n^a n^b},
\end{eqnarray}
where we use as internal variable,
describing internal degrees of freedom at any point $P$, 
the unit timelike vector $n^a$
(instead of a null vector or a unit vector in Euclideanised space
as in \cite{Pad00})  
and the last step is in the limit
$
L^2 E \ll 1. 
$
Here the products extend to all subsystems, 
labelled by pairs $(n, x)$, 
in which 
the phase space generated by
vectors $n_a$ and spacetime coordinates $x^a$
is assumed to be subdivided;
the phase space is constructed with $n_a$ of constant norm ($ = -1$),
and each subsystem is $3\times 4 $ -dimensional.
The quantity $S$ is supposed
to denote at this stage some sort of entropy
describing, in the microcanonical ensemble,
the statistical system, assumed to be isolated,
composed of both matter and spacetime
degrees of freedom;
we will learn something more on it later on.
We see that $\rho_m$ also, as well as $\rho$,
does depend on the discretization scale $L$.
$T_{ab} n^a n^b$ is the mass-energy density of matter
according to an observer with 4-velocity $n^a$.
$L^3 T_{ab} n^a n^b$ is the mass-energy of matter
rescaled to the elemental volume
$L^3$ set by the discretization procedure.
$L^4 T_{ab} n^a n^b$ finds then some meaning as entropy:
$L^4 T_{ab} n^a n^b \equiv (S_m)_{n, x}$
could be interpreted
as entropy, rescaled to a volume $L^3$, that the matter would have,
were the given mass-energy integrally heat
at temperature $T = \frac{1}{L}$.
In other words,
it would be the matter entropy,
rescaled to the elemental
discretized volume (`elemental area' of hypersurfaces) $L^3$,
as `measured' by constituent particles that fit in that volume
(i.e. the entropy we would have if that given mass-energy in $L^3$
were due to constituent particles which fit in a box of edge $L$). 
$ 
\rho_{m}(n, x) 
$
represents formally the associated statistical weight
$\Delta\Gamma_m(n, x) = e^{(S_m)_{n, x}}$,
$\sum_{n, x} (S_m)_{n, x} = S_m$, for subsystem $(n, x)$ for matter,
meaning the number of microscopic states 
for subsystem $(n, x)$ 
associated to the assigned macroscopic state
in flat spacetime \cite{LandA}.
%

Formula (\ref{expS}) presupposes
statistical independence among subsystems,
and assigns to $\rho \rho_m$ the meaning
of something proportional to the total statistical weight
of the macroscopic state (of subsystem $(n, x)$),
i.e. proportional to the total number
(matter + spacetime) of microscopic states
corresponding to the assigned macroscopic configuration.
Indeed,
defining, along what we just did for matter,
$\Delta\Gamma(n, x)$ ($\Delta\Gamma_{flat}(n, x)$) as
the number of microscopic states of spacetime
in subsystem $(n, x)$ which are associated to
the actual spacetime macroscopic configuration
(to the particular configuration provided by flat spacetime),
and putting
$r(n, x) \equiv
\frac{\Delta\Gamma(n, x)}{\Delta\Gamma_{flat}(n, x)}$,
we have

\begin{eqnarray}\label{ratio}
e^S
&=& 
\prod_{n} \ \prod_{x} \ \Delta\Gamma(n, x) \ \Delta\Gamma_m(n, x) \nonumber \\
&=&
\prod_{n} \ \prod_{x} \ \Delta\Gamma_{flat}(n, x)
\  r(n, x) \ \Delta\Gamma_m(n, x)  \nonumber \\
&=&
N \ \prod_{n} \ \prod_{x}
\  r(n, x) \ \Delta\Gamma_m(n, x)  \nonumber \\
&=&
N \ \prod_{n} \ \prod_{x} \ r(n, x) \ \rho_m(n, x) \nonumber \\
&=&
M \ \prod_{n} \ \prod_{x} \ \rho(n, x) \ \rho_m(n, x),
\end{eqnarray}
where
$
N = \prod_{n} \prod_{x} \Delta\Gamma_{flat}(n, x)
$
and $M$ are constants
independent both of $(n, x)$
and of the actual configuration,
and the last equality comes from the 1st relation in (\ref{expS}).
In this formula,
we see the role of $\rho$ is that of
ratio of numbers of states
(between the actual configuration
and that provided by flat spacetime),
not that of ratio of numbers of degrees of freedom.
We have

\begin{eqnarray}
\rho(n, x) 
&=&
\frac{N}{M} \ r(n, x)   \nonumber \\
&\equiv&
C \ r(n, x),
\end{eqnarray}
and,
with $C = 1$,
the normalization constant $M$
has the value
$
M = N = \prod_{n} \prod_{x} \Delta\Gamma_{flat}(n, x).
$

Taking $C=1$
(or, equivalently, defining a new $\rho$
which is $1/C$ times the old one, i.e. we redefine
$\rho$ in order for it to coincide with the ratio $r$),
and denoting
$e^{S_{flat}}$
the total number (matter + spacetime) of microscopic states
for the flat spacetime configuration
($
e^{S_{flat}} =
\prod_{n} \prod_{x} \Delta\Gamma_{flat}(n, x) \Delta\Gamma_m(n, x)
$),
equation (\ref{ratio}) can be recast as

\begin{eqnarray}\label{old_34}
e^S 
&=& 
\prod_{n} \ \prod_{x} \ \Delta\Gamma_{flat}(n, x) \
\rho(n, x) \  \Delta\Gamma_m(n, x)   \nonumber \\
&=&
e^{S_{flat}} \ \prod_{n} \ \prod_{x} \ \rho(n, x)    \nonumber \\
&=&
e^{S_{flat}} \ e^{\sum_{n} \sum_{x} \ln\rho(n, x)} \nonumber \\ 
&=&
e^{S_{flat} + \int d^4v \ d^4x \ \delta_D(v^a v_a + 1) \ \ln\rho}, 
\end{eqnarray}
or

\begin{eqnarray}\label{old_35}
S =
S_{flat} + \int d^4v \ d^4x \ \delta_D(v^a v_a + 1) \ \ln\rho,
\end{eqnarray}
where the integral
comes from going to the continuous description:
$
\sum_{x} \rightarrow \int d^4x,
$
where $d^4x$ denotes the covariant volume measure
on spacetime manifold $M$;
$
\sum_{n}
$
with the internal variable $n^a n_a = -1$
gets mapped to
$\int d^4v \ \delta_D(v^a v_a + 1)$
with internal variable $v^a$,
where $d^4v$ is the invariant volume measure
in $4$-dimensional linear space $T_xM$ 
tangent to $M$ at $x$,
and $\delta_D$ is Dirac's $\delta$;
thus
$\sum_{n} \sum_{x}$ with $n^a n_a = -1$
$\rightarrow \int d^4v \ d^4x \ \delta_D(v^a v_a + 1)$,
where
$\int d^4v \ d^4x$ is the covariant volume measure
in the tangent bundle of $M$.
Relations (\ref{old_34}-\ref{old_35}) 
exhibit a 3-fold connection 
among $S$, $S_{flat}$ and a \ $\ln\rho$ term,  
with the latter 
consistently representing numbers of degrees of freedom:
the $\ln\rho$ term represents the difference
between the number of spacetime degrees of freedom
in the actual configuration and the number
of spacetime degrees of freedom we would have
were the spacetime flat.

From (\ref{rho}), (\ref{Delta_expansion})
we see that,
as long as the spacetime obeys the timelike convergence condition
($R_{ab} v^a v^b \ge 0, \forall v^a$ timelike),
$\Delta(p, P) \ge 1$
(from (\ref{Delta_expansion}), this happens also
if the spacetime obeys the spacelike convergence condition
($R_{ab} v^a v^b \ge 0, \forall v^a$ spacelike);
cf. \cite{VisA})
and $\ln\rho$ is non-positive,
so that the term
$
\int d^4v \ d^4x \ \delta_D(v^a v_a + 1) \ln\rho
\equiv - S_K
\ (S_K \ge 0),
$
can be interpreted as
an absence of degrees of freedom;
i.e. the generic configuration
has less degrees of freedom
than when the spacetime is flat.
Thanks to (\ref{mu_L})
these degrees of freedom can more precisely be thought of
as building blocks of extrinsic curvature $K$,
i.e. they are the degrees of freedom which must be erased
when building up the latter. 
Rewriting $e^S$ as

\begin{eqnarray}\label{expS_3}
e^S
&=&
N \ e^{S_m + \int d^4v \ d^4x \ \delta_D(v^a v_a + 1) \ln\rho},
\end{eqnarray}
we get

\begin{eqnarray}
S = \ln N + S_m -S_K,
\end{eqnarray}
which tells us that
the total number of degrees of freedom ($S$)
is
given
by those due to matter ($S_m$)
plus those of flat spacetime ($+\ln N$)
minus those erased from building up curvature ($-S_K$).
From (\ref{expS_3}) we get

\begin{eqnarray}
e^S
&=&
N \ e^{\int d^4v \ d^4x \ \delta_D(v^a v_a + 1) \ [\ln\rho_m(v, x) + \ln\rho(v, x)]} \nonumber \\
&=&
e^{\int d^4v \ d^4x \ \delta_D(v^a v_a + 1) \ln\rho_{flat}} \
e^{\int d^4v \ d^4x \ \delta_D(v^a v_a + 1) \ [\ln\rho_m(v, x) + \ln\rho(v, x)]} \nonumber \\
&=&
N e^{\int d^4v \ {\cal F}(v^a)},
\end{eqnarray}
with $\ln\rho_{flat}$ independent of $v$ and $x$
due to the isotropy and homogeneity of the flat spacetime,
and
${\cal F}(v^a) \equiv
\int d^4x \delta_D(v^a v_a +1) [\ln\rho_m(v, x) + \ln\rho(v, x)]$.

Proceeding analogously to \cite{Pad00},
we assume now that field equations are what arise
from the extremisation of the quantity
${\cal F}(v^a)$
with respect to the internal variable $v^a$ for each $v^a$,
i.e. from
extremising
${\cal X}(v^a) \equiv \int d^4x [\ln\rho_m(v, x) + \ln\rho(v, x)]$
subject to the constraint $v^a v_a = -1$.
To require this extremisation is equivalent to requiring
that
$
  \int d^4x [\ln\rho_m(v, x) + \ln\rho(v, x)]
$
be constant,
i.e. independent of $v^a$,
when $v^a$ is constrained to be unit timelike.
From the arbitrariness of the integration volume,
this means to require

\begin{eqnarray}
  \ln\rho_m(v, x) + \ln\rho(v, x) = \lambda(x),
  \nonumber
\end{eqnarray}
with $\lambda(x)$ generic,
or,
in the limit
$L^2 R_{ab} v^a v^b \ll 1$,

\begin{eqnarray}
  L^4 T_{ab} v^a v^b - \frac{1}{6} L^2 R_{ab} v^a v^b
  &=& \lambda(x)
  \nonumber \\
  &=& - \lambda(x) \ g_{ab} v^a v^b
  \nonumber
\end{eqnarray}
for any $v^a$ unit timelike,
which means

\begin{eqnarray}
  L^4 T_{ab} - \frac{1}{6} L^2 R_{ab} + \lambda(x) \ g_{ab} = 0.
  \nonumber
\end{eqnarray}  
From this,
using of
$\nabla_a {G^a}_b = 0 = \nabla_a {T^a}_b$
with $G_{ab}$ the Einstein tensor,
we get
$
\partial_a[-\frac{1}{12} L^2 R + \lambda(x)] = 0,
$
and thus
$
-\frac{1}{12} L^2 R + \lambda(x) = {\rm const} \equiv -\frac{1}{6} L^2 \Lambda,
$
with $\Lambda$ constant
(going to play the role of cosmological constant) \cite{PadZ}.
The equation above is then

\begin{eqnarray}
  G_{ab} + \Lambda g_{ab} = 6 L^2 T_{ab}.
  \nonumber
\end{eqnarray}
On choosing $L = {\bar L}$ such that

\begin{eqnarray}
  {\bar L}^2 = \frac{4 \pi}{3} L_{Pl}^2
\end{eqnarray}
with $L_{Pl}$ the Planck length,
we get

\begin{eqnarray}
  G_{ab} + \Lambda g_{ab} = 8 \pi  L_{Pl}^2 T_{ab}.
  \nonumber
\end{eqnarray}
This just confirms (in the case of timelike internal variable) 
that Einstein's equations + cosmological constant 
are what emerges
from the $\rho$-based statistical approach to gravity
in the appropriate limit
(generically, $L_{Pl}^2 |{\cal R}| \ll 1$, where
${\cal R}$ is a typical component of the Ricci tensor) 
with a suitable choice of minimal length scale $L$,
with the cosmological constant arising as an
integration constant \cite{PadX, PadY, PadZ, Pad00}.


\end{document}